\documentclass[conference]{IEEEtran}
\IEEEoverridecommandlockouts
\usepackage{cite}
\usepackage{amsmath,amssymb,amsfonts}
\usepackage{algorithmic}
\usepackage{graphicx}
\usepackage{caption}
\usepackage{textcomp}
\usepackage{subcaption}
\usepackage{xcolor}

\newtheorem{definition}{Definition}

\def\BibTeX{{\rm B\kern-.05em{\sc i\kern-.025em b}\kern-.08em
    T\kern-.1667em\lower.7ex\hbox{E}\kern-.125emX}}
\begin{document}

\newcommand{\norm}[1]{\left\lVert#1\right\rVert}
\title{UAV-Aided Decentralized Learning over Mesh Networks
\thanks{The work of M. Zecchin is funded by the Marie Sklodowska Curie action WINDMILL (grant No. 813999).}
}

\author{
    \IEEEauthorblockN{Matteo Zecchin, David Gesbert and Marios Kountouris}
    \IEEEauthorblockA{Communication Systems Department\\
    EURECOM, Sophia Antipolis, France
    \\\{matteo.zecchin, david.gesbert, marios.kountouris\}@eurecom.fr}
}

\maketitle

\begin{abstract}
Decentralized learning empowers wireless network devices to collaboratively train a machine learning (ML) model relying solely on device-to-device (D2D) communication. It is known that the convergence speed of decentralized optimization algorithms severely depends on the degree of the network connectivity, with denser network topologies leading to shorter convergence time. Consequently, the local connectivity of real world mesh networks, due to the limited communication range of its wireless nodes, undermines the efficiency of decentralized learning protocols, rendering them potentially impracticable. In this work we investigate the role of an unmanned aerial vehicle (UAV), used as flying relay, in facilitating decentralized learning procedures in such challenging conditions. We propose an optimized UAV trajectory, that is defined as a sequence of waypoints that the UAV visits sequentially in order to transfer intelligence across sparsely connected group of users. We then provide a series of experiments highlighting the essential role of UAVs in the context of decentralized learning over mesh networks.
  
\end{abstract}

\begin{IEEEkeywords}
Decentralized learning, unmanned aerial vehicles, machine learning, mesh networks.
\end{IEEEkeywords}

\section{Introduction}
Decentralized learning is a promising distributed optimization paradigm that enables a network of wireless edge devices to collaboratively train a machine learning (ML) model, and to harness the aggregate computational and data resources of the network. Differently from federated learning (FL) \cite{mcmahan2017communication}, decentralized training is not limited to star network topologies with a central orchestrator, but is more flexible as it allows wireless edge devices to exchange training related information in a peer-to-peer (server-less) manner \cite{tsitsiklis1986distributed}. For this reason, they are particularly appealing for future wireless networks based on device-to-device (D2D) communication.

Many different decentralized learning schemes over wireless networks have been recently proposed and analyzed \cite{xing2020decentralized, ozfatura2020decentralized, Xing21-Simeone,Shi21,jeong2022asynchronous}.
The major assumption in all these works is that the network topology is strongly connected on average. However, real world mesh networks are characterized by local, rather than global connectivity, and groups of nodes are often isolated or sparsely connected to the rest of the network due to their limited communication range. In these scenarios, decentralized learning is either not possible or its performance is severely hampered.

At the same time, unmanned aerial vehicles (UAVs) represent an appealing solution to mitigate limited ground connectivity. UAVs have been used as smart flying relays to improve multi-hop routing capabilities \cite{esrafilian2020autonomous}, to self-organize in flying mesh networks \cite{behnke2011comparison} and to improve coverage to ground users \cite{sabino2018topology}. In this work, we investigate the role of UAVs in aiding decentralized learning protocols over ground mesh wireless networks.

The combination of FL and UAV assisted communication has recently been explored; however, these studies have been limited to scenarios in which the UAV has the role of a parameter server (PS), i.e., aggregating model estimates received from ground nodes and subsequently broadcasting the aggregated model back to the ground \cite{donevski2021federated,mrad2021federated}.

Our work differs from these previous works as it considers the UAV serving as a relay and it assumes that ground nodes are able to carry out learning even in the absence of a UAV, by exploiting the already existing ground D2D links. This feature dramatically improves the convergence speed, versatility and fault tolerance of the proposed solution. We propose an optimized trajectory, given as a sequence of waypoints visited by the UAV, which is designed to intelligently provide relaying opportunities to ground nodes and to diffuse locally optimized model across subsets of users with limited connectivity. We provide experiments showing that with the aid of a UAV following the proposed trajectory, it is possible to harness the full potential of the mesh network in spite of sparse and local connectivity, and to accelerate learning compared to UAV-aided federated learning algorithms.

\section{System Model}
We consider a network of $m+1$ devices comprising $m$ ground users plus a UAV serving as a flying relay. We index the ground user by $1,\dots,m$ and we denote the location of the $i$-th ground device by $\mathbf{p}_i=[x_i,y_i,z_i]\in\mathbb{R}^3$, where $x_i$ and $y_i$ are the horizontal coordinates while $z_i$ denotes the elevation. We assume that ground devices are \emph{static}, namely their position is not a function of time. On the other hand, the UAV location is denoted $\mathbf{p}_{uav}=[x,y,z]\in\mathbb{R}^3$ and is assumed to be \emph{time-varying} in the horizontal coordinates $x$ and $y$, but not in the vertical one $z$.  Furthermore, the UAV elevation $z$ is set to be larger than a safety altitude $z_{min}$.  

\subsection{Communication Model}
 Communication among network nodes takes place in rounds. At every communication round $t\in\{\tau,2\tau,\dots\}$, the channel gain coefficient  $g^{(t)}_{i,j}\in\mathbb{R}$, expressed in dB, between each pair of distinct user $(i,j)\in[1:m]^2$ is given by
 \begin{align}
 g^{(t)}_{i,j}=g^{(t)}_{j,i}=\beta_{g}-\alpha_{g}10\log_{10}d_{i,j}^{}+\eta_g^{(t)}
 \end{align}
 where $\alpha_{g}$ is the path loss exponent, $\beta_g$ is the average channel gain in dB at a reference distance $d=1$, $d_{i,j}=\norm{\mathbf{p}_i-\mathbf{p}_j}_2$ is the distance between nodes $i$ and $j$, and $\eta_g\sim\mathcal{N}(0,\sigma^2_g)$ models the shadowing effects. For simplicity, we assume that the link parameters $\alpha_g, \beta_g$ and $\sigma_g$ are homogeneous across pairs of ground users; however, the proposed solution can easily accommodate heterogeneous channel parameters.
 At communication round $t$, the channel gain link between the UAV and a ground node $i$ under Line-of-Sight (LoS) conditions is modeled as 
  \begin{align}
 g^{(t)}_{i,L}=\beta_{L}-\alpha_{L}10\log_{10}d_{i}^{(t)} +\eta_{L}^{(t)},
  \end{align}
  while under Non-Line-of-Sight (NLoS) propagation it follows
\begin{align}
  g^{(t)}_{i,N}=\beta_{N}-\alpha_{N}10\log_{10}d_{i}^{(t)}+\eta_{N}^{(t)} 
 \end{align}
 where $d_{i}^{(t)}=\norm{\mathbf{p}_i-\mathbf{p}_{uav}^{(t)}}_2$ denotes the time-dependent distance between the UAV and user $i$, $\alpha_{L}, \beta_{L}$ and $\eta^{(t)}_{L}\sim \mathcal{N}(0,\sigma_{L}^2)$ are the channel parameters under LoS, while $\alpha_{N},\beta_{N}$ and $\eta^{(t)}_{N}\sim \mathcal{N}(0,\sigma_{N}^2)$ describe the channel under NLoS propagation. These parameters are assumed to be homogeneous across users for simplicity.
 
The LoS probability between the UAV at a position $\mathbf{p}_{uav}^{(t)}$ and user $i$ is modeled using the $s$-model \cite{al2014optimal}
\begin{align}
  \rho_i^{(t)}=\frac{1}{1+e^{-a_i \theta_i^{(t)}+b_i}}
\end{align}
where $a_i$ and $b_i$ are model coefficients related to the propagation environment, and $\theta^{(t)}_i$ is the elevation angle between the UAV and ground user $i$.
At every communication round $t$, a link between network nodes is modeled using a simple, yet classical, on-off channel model. Accordingly two nodes communicate if and only if the associated channel gain exceeds a threshold $g_{th}$. Therefore, the resulting ground connectivity matrix  $A^{(t)}_{gr}\in[0,1]^{m\times m}$, is symmetric, has diagonal elements being equal to 1 and Bernoulli distributed off-diagonal entries
  \begin{align}
[A^{(t)}_{gr}]_{j,i}=[A^{(t)}_{gr}]_{i,j}\sim Bern\left(1- \Phi\left(\frac{g_{th}-\bar{g}_{i,j}^{(t)}}{\sqrt{\sigma_g}}\right)\right)
\end{align}
where $\Phi(\cdot)$ denotes standard Gaussian cumulative distribution function and $\bar{g}_{i,j}^{(t)}=\mathbb{E}[g_{i,j}^{(t)}]$.

Similarly, the connectivity between the UAV and ground users is described by a vector $\mathbf{a}^{(t)}_{uav}\in[0,1]^{1\times m}$ with entries
\begin{align}
[\mathbf{a}^{(t)}_{uav}]_i\hspace{-0.2em}\sim\hspace{-0.2em} Bern\left(\hspace{-0.2em}1\hspace{-0.2em}-\hspace{-0.2em}\bar{\rho}_i\Phi\hspace{-0.2em}\left(\frac{g_{th}-\bar{g}_{i,N}^{(t)}}{\sqrt{\sigma_{N}}}\right)\hspace{-0.2em}-\hspace{-0.2em}\rho_i\Phi\hspace{-0.2em}\left(\frac{g_{th}-\bar{g}_{i,L}^{(t)}}{\sqrt{\sigma_{L}}}\right)\hspace{-0.3em}\right)
\end{align}
where $\bar{\rho}_i=1-\rho_i$, $\bar{g}_{i,L}^{(t)}=\mathbb{E}[g_{i,L}^{(t)}]$ and $\bar{g}_{i,N}^{(t)}=\mathbb{E}[g_{i,N}^{(t)}]$.

Based on the instantaneous connectivity status, determined by the realization of $\mathbf{a}^{(t)}_{uav}$, the UAV serves as a one-hop relay for the communication among grounds users. The connectivity matrix resulting from the relaying opportunities offered by the UAV to ground users is obtained as
 \begin{align}
A^{(t)}_{uav}&=(\mathbf{a}^{(t)}_{uav})^T\mathbf{a}^{(t)}_{uav}.
\end{align}
If follows that $A^{(t)}_{uav}$ is a symmetric random binary matrix whose entry $(i,j)$ is $1$ if an only if there exists a relaying opportunity between ground user $i$ and $j$, and $0$ otherwise.

Overall, the aggregated connectivity matrix, accounting for link existence either by D2D ground communication or thanks to UAV relaying, is given by
  \begin{align}
A^{(t)}&=J_m-(J_m-A^{(t)}_{uav})\odot(J_m-A^{(t)}_{gr})
\end{align}
 where $J_m$ is the $m\times m$ all-one matrix and $\odot$ denotes the Hadamard product. 
 
 For every realization of the connectivity matrix $A^{(t)}$, the set of devices connected to node $i$ is
 \begin{align}
 \mathcal{N}^{(t)}(i):=\{j : [A^{(t)}]_{i,j}=1\}.
\end{align}
Note that every ground user is connected to itself.
\iffalse
\begin{figure*}[]
\minipage{0.33\textwidth}
  \includegraphics[width=\linewidth]{Images/Traj.pdf}
  \caption{A really Awesome Image}\label{fig:11}
\endminipage\hfill
\minipage{0.33\textwidth}
  \includegraphics[width=\linewidth]{Images/Acc.pdf}
  \caption{A really Awesome Image}\label{fig:12}
\endminipage\hfill
\minipage{0.33\textwidth}%
  \includegraphics[width=\linewidth]{Images/ConsErr.pdf}
  \caption{A really Awesome Image}\label{fig:13}
\endminipage
\end{figure*}
\fi
\subsection{Learning procedure}
We assume that the goal of ground devices is to collaboratively train a machine learning model in order to benefit from the aggregation of local computational resources and in-situ data.
In particular, we assume that each ground device is endowed with a local loss function $f_i:\mathbb{R}^d\to \mathbb{R}$ and local parameter estimate $\theta_i\in \mathbb{R}^d$. This corresponds to the distributed empirical risk minimization problem, which entails most decentralized learning problems, whenever $\{f_i\}^m_{i=1}$ are loss terms defined over local datasets $\{\mathcal{D}_i\}^m_{i=1}$ and $\{\theta_i\}^m_{i=1}$ are local model estimates. 

The network objective consists in the minimization of the aggregate network loss subject to a consensus constraint
\begin{align}
\underset{\theta_1,\dots,\theta_m}{\text{minimize}}\ &f(\theta_1,\dots,\theta_m) := \frac{1}{m} \sum_{i=1}^m f_i(\theta_i)\label{problem} \\
\text{s.t.\quad } &\theta_1=\theta_2=\dots=\theta_m. \nonumber
\end{align}
In the following, we denote the average network estimate as $\bar{\theta}=1/m\sum^m_{i=1}\theta_i$.

To solve (\ref{problem}), we consider the asynchronous decentralized stochastic gradient descent (DSGD) algorithm proposed in \cite{jeong2022asynchronous}. According to this optimization scheme, ground devices alternate between a local optimization phase based on gradient information (computation phase) and a communication phase to exchange the updated local estimates with one-hop neighbours.  To locally optimize the model estimate $\theta_i$, we assume that each device $i$ can query a stochastic oracle that is unbiased,
\begin{align}
    \mathbb{E}[g_i(\theta_i)]=\nabla_\theta f_i(\theta_i),
\end{align}
and has bounded variance and magnitude 
\begin{align}
&\mathbb{E}\norm{ g_i(\theta_i)-\nabla_\theta f_i(\theta_i)}^2\leq \sigma^2,\\
&\mathbb{E}\norm{g_i(\theta_i)}\leq G^2.
\end{align}
Furthermore, to account for computation impairments and energy constraints, we admit the existence of straggling ground users that can become inactive or postpone the local optimization computation.
At each communication round $t$, the local update rule at device $i$ becomes
\begin{equation}
    \theta_i^{(t+\frac{1}{2})}=\begin{cases}
\theta_i^{(t)} , \quad \  \text{ if device $i$ is straggler at round $t$}\\
\theta_i^{(t)} -\eta^{(t)}_ig_i(\theta^{(t-\tau_i)}) \quad \text{otherwise}
\end{cases}
\label{eq:update}
\end{equation}
where $\eta^{(t)}_i$ is a local learning rate and the delay $\tau_i\geq 0$ accounts for the staleness of the gradient information at device $i$. Subsequently, each device $i$ shares its updated local estimate $ \theta_i^{(t+\frac{1}{2})}$ with its neighbours $ \mathcal{N}^{(t)}(i)$ using either a digital or analog communication protocol \cite{Xing21-Simeone,ozfatura2020decentralized}. The received estimates are then averaged to obtain the new local estimate 
\begin{equation}
    \theta_i^{(t+1)}=\sum_{j\in\mathcal{N}(i)}w_{i,j}  \theta_j^{(t+\frac{1}{2})}
    \label{eq:aggregation}
\end{equation}
 where $w_{i,j}$ are the entries of the mixing matrix $W^{(t)}$ obtained using a Metropolis-Hastings weighting scheme \cite{xiao2006distributed}.

In \cite{jeong2022asynchronous}, it has been shown that the performance of the asynchronous DSGD optimization procedure depends both on the activity of users and on the degree of wireless network connectivity. In particular, with more connected network topologies converging faster than sparser ones. This motivates the use of a UAV to facilitate the diffusion of locally optimized models, and to render the decentralized learning protocol more efficient in spite of sparse and local ground connectivity.
\iffalse
\begin{definition}[Expected Consensus Rate \cite{koloskova2020unified}]
A sequence of random matrices ${W^{(t)}}\in\mathbb{R}^{m\times m}$ is said to satisfy the expected consensus with rate $p$ if for any $X\in \mathbb{R}^{d\times m}$, if for any $l>0$
\begin{align}
\mathbb{E}_{W}\left[\norm{{W}_{l,t}X-\bar{X}}^2_F\right]\leq (1-p)\norm{X-\bar{X}}^2_F
\end{align}
where $W_{l,t}=W^{(l+1)\tau-1}\dots W^{(l\tau)}$,  $\bar{X}=X\frac{\mathbf{1}\mathbf{1}^T}{m}$ and the expectation is w.r.t. the random matrices $W^{(l+1)\tau-1},\dots ,W^{(l\tau)}$.
\end{definition}
[Example showing how connectivity hampers convergence in wireless networks]\fi
\section{Trajectory Optimization}
At every optimization round $t$, the connectivity matrix $A^{(t)}$ associated to the network of ground users depends on the UAV location $\mathbf{p}_{uav}^{(t)}$ and it can be enhanced thanks to the the relaying opportunities it provides to ground devices. In the following, we propose to optimize the UAV trajectory during the optimization process so that the distributed learning procedure is facilitated.

A key quantity that is used to measure the information diffusion capabilities of a network is the expected consensus rate \cite{koloskova2020unified}. While this quantity can be used to characterize the rate of convergence of DSGD procedure, it does not provides a tractable optimization objective to derive the UAV trajectory. For this reason we define a more tractable surrogate objective that yields the optimized trajectory as a sequence of waypoints $\{w_i\}^n_{i=1}$. In particular, we assume that the initial way point $w_0$ is equal to the initial UAV location $\mathbf{p}_{uav}^{(0)}$, and that the sequence of waypoints is determined on-the-fly, with the waypoint $w_{i+1}$ being computed when the UAV reaches the location specified by the previous waypoint $w_i$. The waypoints are designed so as to hover the UAV on a position that maximizes the probability of creating a relaying opportunities between users that, up to communication round $t$, have not been able to communicate. To this end, we recursively define an link activity rate matrix $R^{(t)}$ as
\begin{align} 
&R^{(0)}=0\\
&R^{(t+1)}=\gamma R^{(t)}+(1-\gamma)\mathbb{E}[A^{(t)}],
\end{align}
for $\gamma\in (0,1)$.
Denoting by $t_i$ the communication round in which the UAV reaches the waypoint $w_i$, the subsequent waypoint is obtained solving the following optimization problem
\begin{align}
\underset{p_t}{\text{maximize}} \norm{(1-R^{(t_i)})\odot \mathbb{E}[A_{uav}]}_{1}
\label{opt_prob}
\end{align}
where $\norm{\cdot}_1$ denotes the entry-wise 1-norm; namely, the sum of absolute values of the matrix entries.

The optimization problem $(\ref{opt_prob})$ determines the next waypoint so as to maximize the relaying opportunities between pair of users associated to links with a low activity rate. The main challenges in solving $(\ref{opt_prob})$ are the lack of a close form expression for $\Phi(\cdot)$ and the non-convexity of the objective. In order to make the objective differentiable, we approximate $\Phi(\cdot)$ using the sigmoid function
\begin{align}
    S(x):=\frac{1}{1+e^{-\alpha x}}
\end{align} 
where $\alpha$ is a fitting parameter set to $\alpha=-1.702$ as proposed in \cite{lieberman2005introduction}.
This approximation of the objective allows us to employ efficient gradient based solvers to generate the sequence of waypoints.

Nonetheless, the optimization objective (\ref{opt_prob}) remains non-convex. In order to reduce the probability of obtaining a waypoint associated to poor local maxima, we employ gradient descent with restarts. The number of restart points is chosen to meet the UAV computation constraints and the restart points are sampled uniformly at random inside the convex hull determined by ground user locations.

\section{Simulations}
\begin{figure*}[]
\begin{subfigure}[t]{.32\textwidth}
  \includegraphics[width=\linewidth]{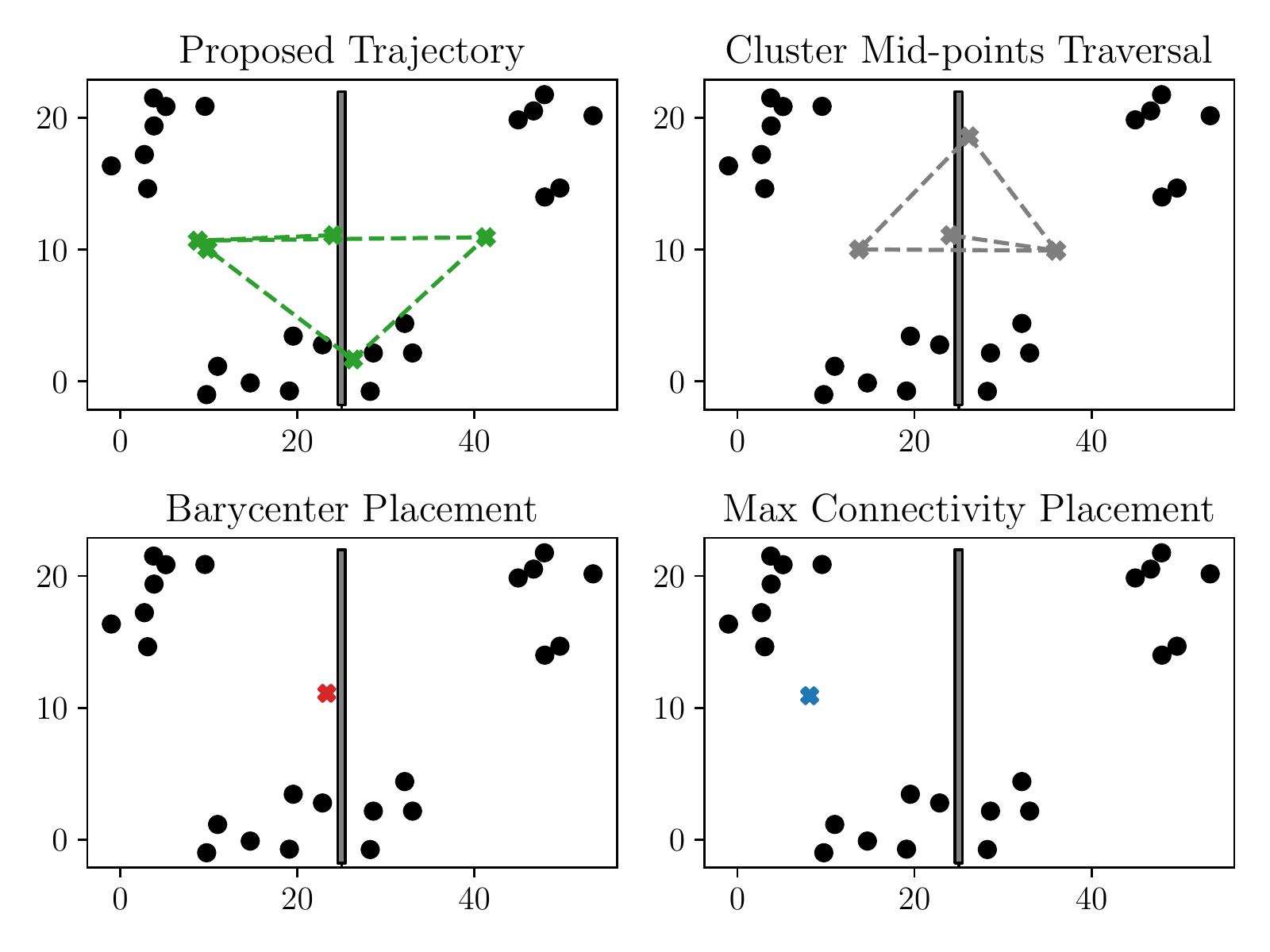}
  \caption{Different UAV trajectory and placements. Black dots represent ground users and the gray vertical line is a propagation obstacle that corresponds to a $35$ dB attenuation.}\label{fig:21}
\end{subfigure}%
\hfill
\begin{subfigure}[t]{.32\textwidth}
  \includegraphics[width=\linewidth]{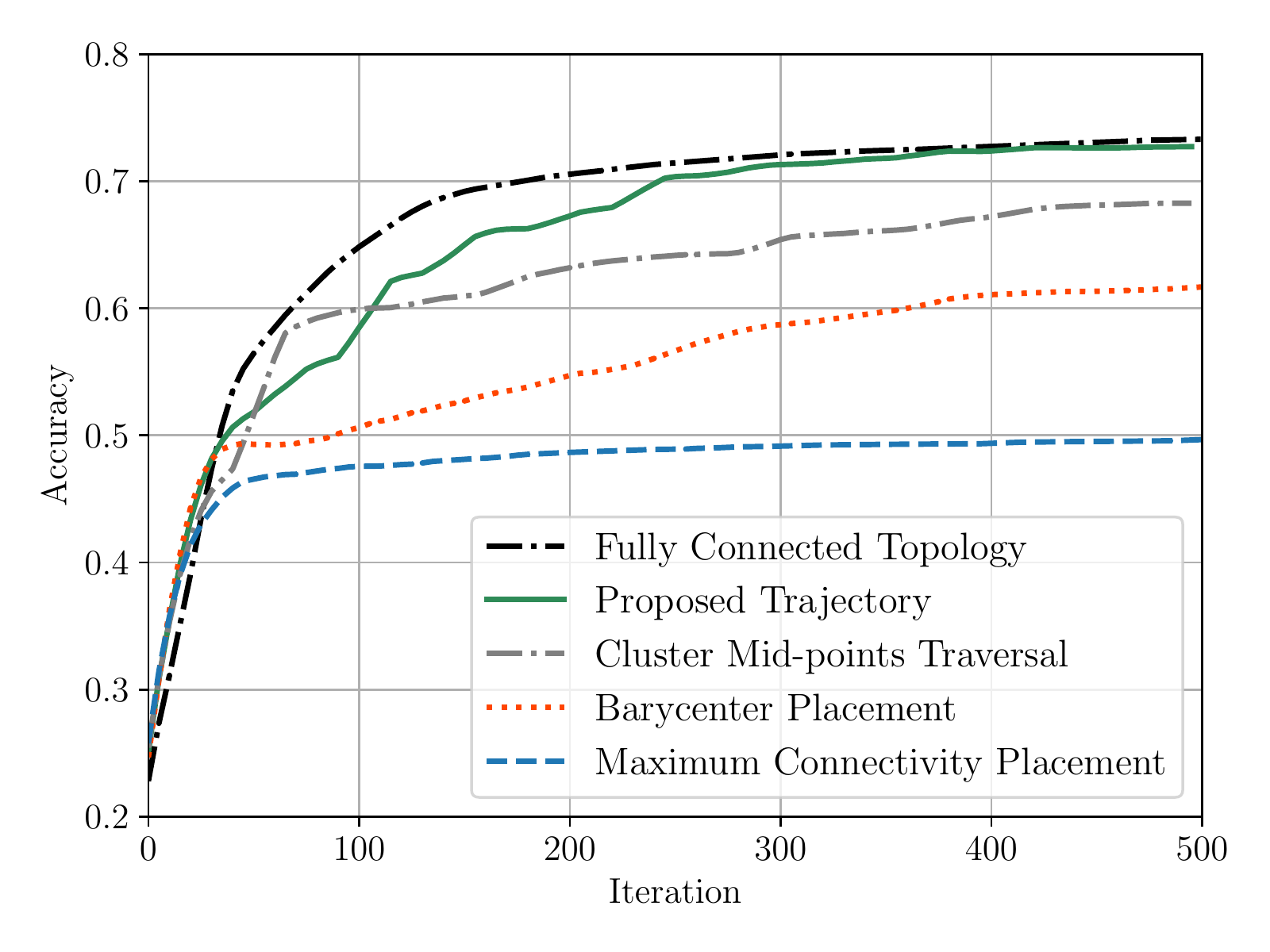}
  \caption{Testing accuracy averaged over 5 runs, obtained by the mean network estimate, using different UAV-Aided decentralized learning protocols.}\label{fig:22}
\end{subfigure}%
\hfill
\begin{subfigure}[t]{.32\textwidth}
  \includegraphics[width=\linewidth]{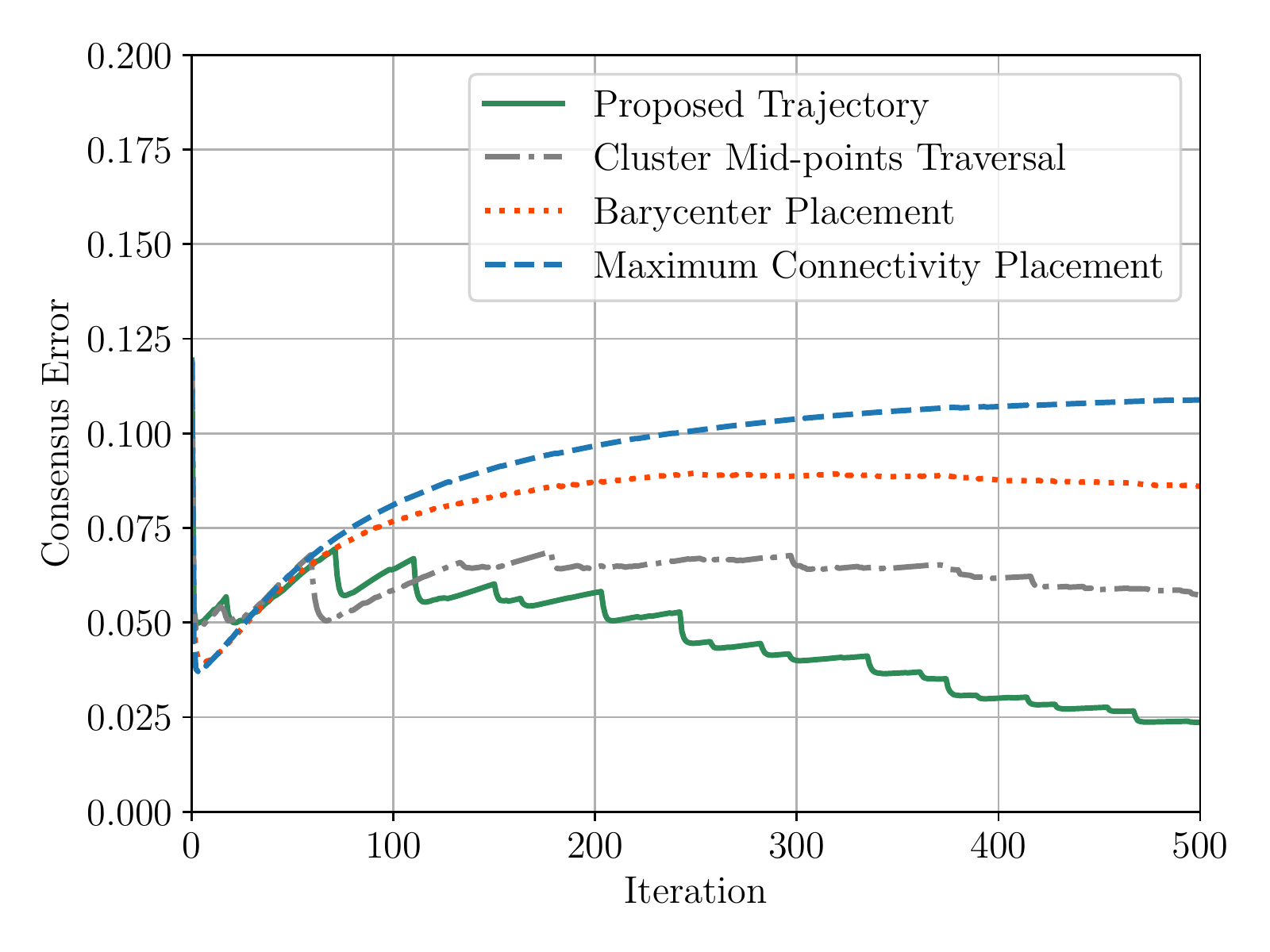}
  \caption{Evolution of the average consensus error (\ref{cons_err}) attained by the benchmarked trajectories. Smaller consensus error corresponds to disagreement between network nodes.}\label{fig:23}
\end{subfigure}
\vspace{-1em}
\end{figure*}

To test the proposed solution, we consider a network deployment of $30\times60\text{m}^2$ with $23$ ground devices deployed at the ground level ($z=0\text{m}$) as depicted in Figure \ref{fig:21}. The propagation parameters describing the ground links channel gain are set to $\alpha_g=3,\beta_g=-30$ dB and $\sigma^2_g=1$, and the channel gain threshold determining active/inactive links is fixed to $g_{th}=-60$ dB. In the considered deployment, ground users are naturally clustered together in 3 distinct groups and the path exponent is such that communication within each cluster is possible, but links between users belonging to different clusters are active with negligible probability. Furthermore, we consider an obstruction  (gray vertical line) that amounts to a $35$ dB attenuation for the links between users residing on opposite side of the line. A UAV flying at a fixed altitude of $10$m serves as a relay to enhance ground connectivity. The channel gain parameters describing the link between ground users and the UAV under LoS propagation are  $\alpha_L=2.5,\beta_L=-30$ dB and $\sigma^2_L=1$, while under NLoS are $\alpha_N=3,\beta_N=-30$ dB and $\sigma^2_N=1$.

We assume that the devices store only $10$ data samples from the FashionMNIST dataset, which alone would not guarantee good inference capabilities. Therefore, they wish to harness the distributed dataset to jointly train a machine learning model. In the following experiments we consider a fully connected neural network with one hidden layers comprising $25$ neurons. Ground devices update the model employing a gradient descent optimizer and a geometrically decaying learning rate $\eta_i^{(t)}=0.1\cdot(0.995)^t$. 

In this setting, the role of the UAV is to intelligently create relaying opportunities so to promote collaborative learning, and to facilitate the ground users to harness the entire distributed dataset by global diffusion of the locally optimized models, in spite of sparse and local connectivity.

We benchmark the proposed solution, corresponding to a UAV visiting the sequence of waypoints returned by (\ref{opt_prob}) and serving users for $20$ optimization rounds each time a waypoint is reached, and compare it with alternative trajectory optimization schemes.

In particular, we consider the \emph{cluster mid-points traversal} trajectory according to which the UAV first runs a $k$-means clustering algorithm to detect natural clusters of the ground devices positions, and then visits sequentially the mid-points between each pair of cluster centers. Once the mid-point is reached, the UAV serves ground users for $20$ optimization rounds and then it flies to the next location.

We consider the \emph{barycenter} placement, in which the UAV hovers at a fixed location $\mathbf{p}^{bar}_{uav}$ determined by the mean ground user location.

Finally, we consider the \emph{maximum connectivity} placement in which the UAV location is fixed and set equal to the coordinate that maximizes the probability of creating a relay links. This is obtained solving the following maximization problem
\begin{align}
\underset{p}{\text{maximize}} \norm{(1-A_{gr})\odot \mathbb{E}[A_{uav}]}_1.
\end{align}
For both the barycenter and the maximum connectivity placements, the UAV serves ground users at every communication round as it hovers at a fixed location for the entire learning procedure. We consider the centralized learning solution, corresponding to a \emph{fully connected} topology, as a performance upper bound.
For all listed approaches we run the distributed optimization protocol for $500$ rounds and we track the testing accuracy of the mean network estimate $\bar{\theta}$. We also consider the network consensus error metric
\begin{align}
    \varepsilon(\theta_1,\dots,\theta_m)=\frac{1}{dm}\sqrt{\sum_{i=1}^m\norm{\theta_i-\bar{\theta}}^2_2},
    \label{cons_err}
\end{align}
which measures the degree of disagreement among ground nodes estimates and it is a proxy to assess how effective is the UAV trajectory in satisfying the consensus constraint in (\ref{opt_prob}).

In Figure \ref{fig:21} we plot the ground user locations (black dots) and we overlay the UAV trajectories during the training process. Specifically, in the top left corner, we report in green the UAV trajectory returned by the proposed scheme. The UAV frequently hovers between the disconnected components in order to relay information between the clusters and to diffuse model estimates between groups of interconnected ground nodes that rarely communicate using D2D ground links. In the top right corner, we provide the trajectory of the cluster mid-points traversal solution (gray). The UAV successfully identifies the clusters and it sequentially visits the mid-points. This strategy enhances the ground connectivity, but it fails at providing relaying opportunities across the two bottom components that are disconnected due to the propagation obstacle. In the bottom row, we plot the barycenter (left plot) and maximum connectivity (right plot) placementes. Both solutions yields a static UAV placement that is fixed throughout the entire training phase.

In Figure \ref{fig:22} we report the testing accuracy attained by the mean network estimate $\bar{\theta}$ when decentralized learning is assisted by a UAV flying according to the different trajectories. The testing accuracy allows us to quantify the extent to which the UAV is beneficial to the collaborative learning process. The proposed solution (in green) is able to take full advantage of the distributed dataset, and it successfully enables fast distributed training with a final accuracy level that matches the accuracy of the centralized solution (in black). The barycenter solution (in red) converges slowly to a lower accuracy level, highlighting the necessity of a dynamic UAV placement to take full advantage of the network resources. The cluster mid-points traversal solution, despite enhancing ground connectivity, it is not able to connect all the disjoint components and therefore it converges to suboptimal solution. Similarly, the maximum connectivity placement is not able to connect all the disjoint network components and to transfer intelligence across different clusters.

In Figure \ref{fig:23} we report the network consensus error evolution attained by the different UAV trajectories. While the proposed trajectory is able to reduce the consensus error during training and it eventually ensures that the edge devices reach a common learning goal, the other baselines are not able to drive network nodes to a globally shared model estimate.

\begin{figure}[]
  \centering
  \includegraphics[width=1\linewidth]{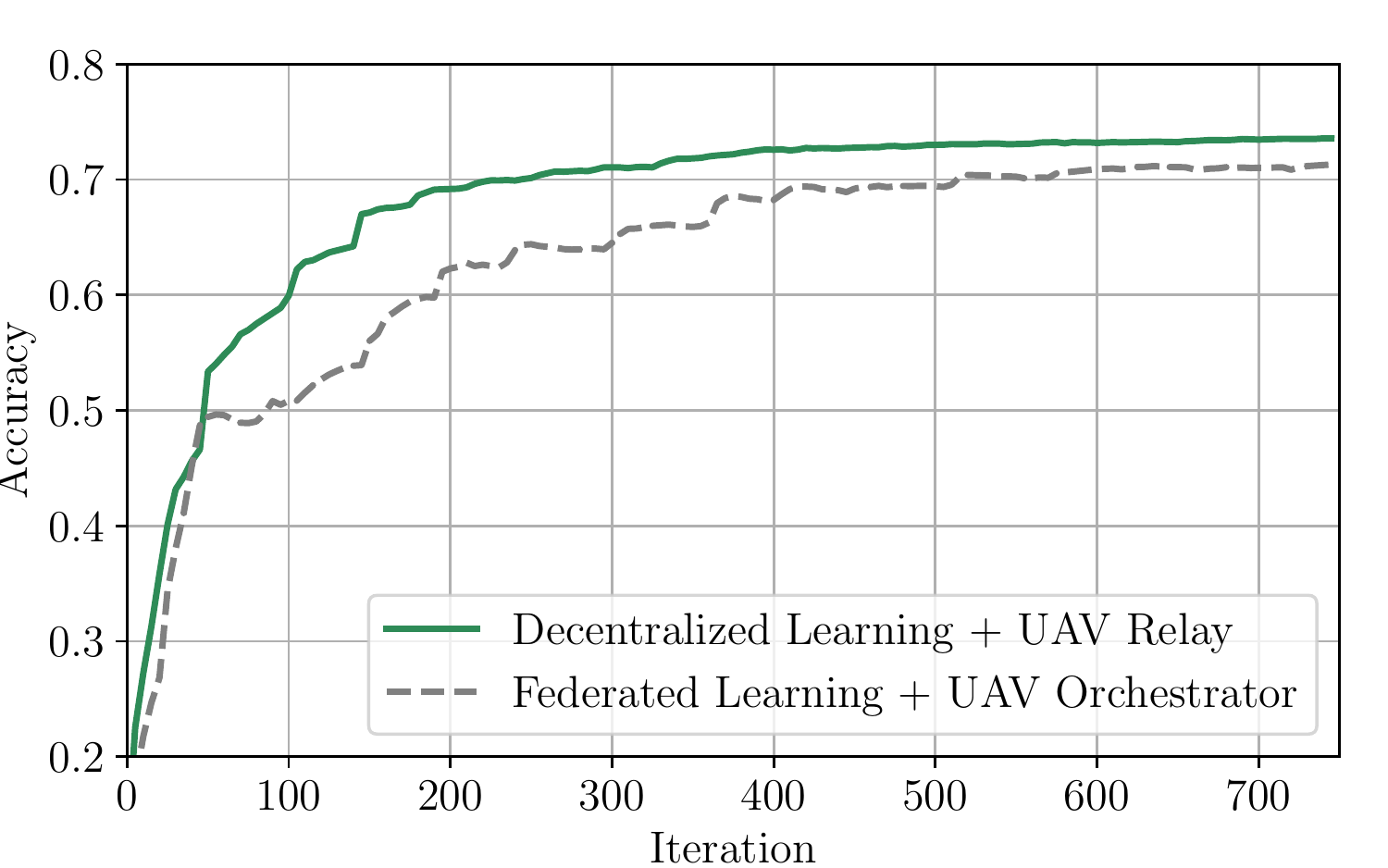}
  \caption{Testing accuracy, averaged over 5 experiments, obtained by the mean network estimate when training is aided by a UAV serving as a relay to assist the decentralized learning protocol (green), or as an orchestrator to perform federated learning (gray dashed).}\label{fig:3}
  \vspace{-1em}
\end{figure}
Finally, we propose a comparison between the proposed decentralized learning scheme and a UAV-aided federated learning protocol, as in \cite{donevski2021federated}. In particular, for the federated learning algorithm, we assume that the UAV serves as a PS, and it orchestrates the learning procedure by collecting locally optimized models by the network devices and broadcasting aggregated estimates back to the ground users. On the other hand, in case of decentralized learning, the UAV serves as a relay and the ground devices can also exploit the available D2D ground links to exchange model estimates, in principle being able to perform learning without the presence of a UAV. As a result, the proposed protocol is more flexible with respect to the communication topology, it can easily accommodate multiple assisting UAVs, and converges faster. To compare these two approaches we study the same deployment as in Figure \ref{fig:21}. We assume that the relaying UAV follows the proposed trajectory, while the UAV serving as orchestrator follows the trajectory obtained solving (\ref{opt_prob}) setting $A^{(t)}_{gr}=0$, trying to serve large groups of users prioritizing stale ones, akin to \cite{donevski2021federated}. In Figure \ref{fig:3} we report the testing accuracies attained by the protocols. The proposed approach drastically reduces the training time, halving the number of iterations required to reach the final performance obtained by the federated learning protocol.

\section{Conclusion}
In this work, we studied the benefits that a flying relay can bring to a network of wireless devices that are jointly training a machine learning model. We proposed a trajectory optimization scheme that enhances the ground connectivity so as to facilitate the diffusion of locally optimized model estimates, and that enables ground users to take full advantage of network computational and data resources. We provided a series of experiments highlighting how a properly designed UAV trajectory can greatly promote decentralized training and outperform UAV-aided federated learning protocols.

\bibliographystyle{ieeetr}  % appearance order
\bibliography{refs}

\end{document}